\documentclass[a4paper,11pt]{article}

\pdfoutput=1
\usepackage{amsmath}
\usepackage{graphicx}

\usepackage[english]{babel}
\usepackage[utf8]{inputenc}
\usepackage{pdflscape}
\usepackage{enumerate}
\usepackage{amsbsy}
\usepackage{amsmath} 
\usepackage{graphics}
\usepackage{mathrsfs}
\usepackage{wrapfig}
\usepackage{mathtools}

\usepackage{amsfonts}
\usepackage{pstricks}
\usepackage{color}
\usepackage{setspace}

\usepackage{accents}
\usepackage{tensor}

\usepackage{jcappub_ver} 

\newcommand{\bea}{\begin{eqnarray}} \newcommand{\eea}{\end{eqnarray}}
\newcommand{\el}{\nonumber \\}
\newcommand{\re}[1]{(\ref{#1})}

\newcommand{\pat}{\partial}

\renewcommand{\sec}[1]{section \ref{#1}}

\renewcommand{\a}{\alpha}
\renewcommand{\b}{\beta}
\renewcommand{\c}{\gamma}
\renewcommand{\d}{\delta}
\newcommand{\e}{\epsilon}
\newcommand{\m}{\mu}
\newcommand{\n}{\nu}
\newcommand{\s}{\sigma}

\newcommand{\ha}{\frac{1}{2}}

\newcommand{\rmd}{\mathrm{d}}

\newcommand{\ie}{i.e.\ }

\newcommand{\cd}{\nabla}

\newcommand{\lcd}{\mathring{\cd}}

\title{Inflation with the Gauss--Bonnet term in the Palatini formulation}

\author{Ali Hassan}
\author{and Syksy R\"{a}s\"{a}nen}

\affiliation{University of Helsinki, Department of Physics and Helsinki Institute of Physics \\ P.O. Box 64, FIN-00014 University of Helsinki, Finland}

\emailAdd{ali.hassan@helsinki.fi}
\emailAdd{syksy.rasanen@iki.fi}

\abstract{
We consider the Gauss--Bonnet term coupled to the inflaton in the Palatini formulation of gravity. Unlike in the metric formulation, the Gauss--Bonnet term is not always a total derivative. We solve for the connection and insert it into the action, exactly for the spatially flat FLRW spacetime, and using the gradient approximation and order reduction for a general spacetime. We consider three cases: when the connection is unconstrained, and when non-metricity or torsion is put to zero. In all cases, the leading order change to the inflaton kinetic has the same form as that generated by the Chern--Simons term, but with a negative sign. The modification of the gravitational wave sector also has the same form as in the Chern--Simons case but with a negative sign, except possibly for zero torsion, depending on the coupling and the potential. Within the range of validity of our approximations, differences from the metric formulation are small unless the kinetic term flips sign or is close to doing so.
}

\begin{document}

\begin{flushleft}
	\hfill		 HIP-2026-5/TH \\
\end{flushleft}
 
\setcounter{tocdepth}{3}

\setcounter{secnumdepth}{3}

\maketitle

\section{Introduction}

Inflation is the most successful scenario for the early universe \cite{Starobinsky:1979ty, Starobinsky:1980te, Kazanas:1980tx, Guth:1981, Sato:1981, Mukhanov:1981xt, Linde:1981mu, Albrecht:1982wi, Hawking:1981fz, Chibisov:1982nx, Hawking:1982cz, Guth:1982ec, Starobinsky:1982ee, Sasaki:1986hm, Mukhanov:1988jd}, with predictions in excellent agreement with observations \cite{Akrami:2018odb, BICEP:2021xfz, McDonough:2025lzo}. The simplest candidate to drive inflation is a single scalar field, called the inflaton. The inflaton is expected to couple non-minimally to the curvature based on effective field theory arguments \cite{Weinberg:2008hq, Solomon:2017nlh} and explicit loop calculations \cite{Callan:1970ze}. This includes coupling of the field to the Ricci scalar, most prominently discussed in Higgs inflation \cite{Bezrukov:2007, Bauer:2008}, as well as more complicated couplings \cite{Capozziello:1999uwa, Capozziello:1999xt, Daniel:2007kk, Sushkov:2009hk, Germani:2010gm, Germani:2010ux, Kobayashi:2010cm, Kamada:2010qe, Kobayashi:2011nu, Germani:2011mx, Tsujikawa:2012mk, Kamada:2012se, Kamada:2013bia, Germani:2014hqa, Yang:2015pga, Kunimitsu:2015faa, DiVita:2015bha, Escriva:2016cwl, Fumagalli:2017cdo, Sato:2017qau, Fu:2019ttf, Granda:2019wyi, Granda:2019wip, Sato:2020ghj, Fumagalli:2020ody, Gialamas:2020vto, Dioguardi:2023jwa, Nezhad:2023dys, BouzariNezhad:2025bgx}.

Particularly interesting are couplings to the Chern--Simons term (also called the Pontryagin term) \cite{Lue:1998mq, Jackiw:2003pm, Alexander:2009tp} and the Gauss--Bonnet term (also called the Euler term) \cite{Lanczos:1938sf, Lovelock:1970zsf, Lovelock:1971yv, Lovelock:1972vz}, both quadratic in the Riemann tensor. The Chern--Simons term violates parity, the Gauss--Bonnet term is parity-invariant. The Gauss--Bonnet term is part of the series of Lovelock terms, which lead to second order equations of motion in the metric formulation of general relativity. In four spacetime dimensions, it is the only non-zero Lovelock term beyond the cosmological constant and the Ricci scalar. In the metric formulation, both the Chern--Simons term and the Gauss--Bonnet term are total derivatives, so their actions are topological (the Gauss--Bonnet action gives the Euler number of the manifold) and do not contribute to the classical equations of motion. When directly coupled to a scalar field, they become dynamical. The Chern--Simons term is then likely unstable, whereas the Gauss--Bonnet term is likely stable when the scalar field has a kinetic term \cite{Delsate:2014hba, Crisostomi:2017ugk}.

The situation is different in the Palatini formulation (also called the metric-affine formulation) of general relativity, where the metric and the connection are independent degrees of freedom.\footnote{In the Palatini formulation there are also dimension 2 geometrical terms beyond the Ricci scalar, such as the Holst term and the Nieh--Yan term \cite{Nieh:1981, Holst:1995pc} that become dynamical when coupled to a scalar field and can contribute to inflation \cite{Langvik:2020nrs, Shaposhnikov:2020gts}.} The Chern--Simons term remains a total derivative. In contrast, while according to the Chern--Gauss--Bonnet theorem the Gauss--Bonnet term is a total derivative for any metric-compatible connection, this is not always true when non-metricity is non-zero \cite{Nieh:1979hf, Babourova_1997, Nieh:2007zz, Borunda:2008, Nieh:2008btw, Jim_nez_2014, Nieh:2018rlg, Janssen:2019doc, Janssen:2019uao, Iosifidis_2021, Banados:2025sww}. In the metric formulation, the Chern--Simons term does not affect the evolution of the metric in the Friedmann--Lema\^{\i}tre--Robertson--Walker (FLRW) spacetime even when directly coupled to a scalar field, unlike the Gauss--Bonnet term. In the Palatini formulation both change the evolution of the FLRW metric. In the Palatini formulation the Gauss--Bonnet term can introduce new degrees of freedom \cite{Banados:2025sww} some of which may be unstable, like the Chern--Simons term.

In \cite{Hassan:2024vjg}, we studied the Chern--Simons term coupled to the inflaton in the Palatini formulation. We now consider the Gauss--Bonnet term coupled to the inflaton in the Palatini formulation. Inflation with the Gauss--Bonnet term has been much studied in the metric formulation \cite{Wetterich:1989hf, Carter:2005fu, Satoh_2008, vandeBruck:2015gjd, Rashidi_2020, Mudrunka:2025xcg}, but rarely in the Palatini formulation \cite{Gasperini:1992sq, Ozer:2016oex} (see \cite{TerenteDiaz:2023kgc} for an application to quintessence). We consider three cases: when the connection is unconstrained, and when either non-metricity or torsion is put to zero a priori. In \sec{sec:dof} we outline the geometrical setup and give the action and the equations of motion. In \sec{sec:sol} we solve the equation of motion for the connection and insert it back into the action to write the theory only in terms of the metric and the inflaton. For the spatially flat FLRW spacetime, we do this exactly. For a general spacetime, we use the gradient approximation and apply order reduction to eliminate spurious degrees of freedom, and then consider perturbations around the spatially flat FLRW metric. In \sec{sec:conc} we summarise our findings and mention open problems.

\section{Gauss--Bonnet term coupled to a scalar field} \label{sec:dof}

\subsection{Geometrical quantities}

\subsubsection{Nonmetricity, torsion, and curvature}

In the Palatini formulation the metric $g_{\a\b}$ and the connection $\Gamma^\c_{\a\b}$ are independent variables. We decompose the connection, defined with the covariant derivative as $\nabla_\b A^\a=\pat_\b A^\a + \Gamma^\a_{\b\c} A^\c$, $\nabla_\b A_\a=\pat_\b A_\a - \Gamma^\c_{\b\a} A_\c$, as
\bea \label{Gamma}
  \Gamma^\c_{\a\b} &=& \mathring\Gamma^\c_{\a\b} + L^\c{}_{\a\b} = \mathring\Gamma^\c_{\a\b} + J^\c{}_{\a\b} + K^\c{}_{\a\b} \ ,
\eea
where $\mathring\Gamma_{\a\b}^\c$ is the Levi--Civita connection of the metric $g_{\a\b}$ and $L^\c{}_{\a\b}$ is the distortion tensor. We denote quantities defined with the Levi--Civita connection with $\mathring{}$. In the second equality we have decomposed $L_{\a\b\c}$ into the disformation $J_{\a\b\c}$ and the contortion $K_{\a\b\c}$, defined as
\bea \label{JK}
  J_{\a\b\c} &\equiv& \frac{1}{2} \left(Q_{\a\b\c}  - Q_{\c\a\b} - Q_{\b\a\c} \right) \ , \quad K_{\a\b\c} \equiv \ha (T_{\a\b\c} + T_{\c\a\b} + T_{\b\a\c} ) \ ,
\eea
where $Q_{\a\b\c}$ and $T_{\a\b\c}$ are the non-metricity and the torsion, respectively, defined as
 \bea \label{QT}
   Q_{\a\b\c} \equiv \nabla_\a g_{\b\c} = - 2 L_{(\c|\a|\b)} = - 2 J_{(\c|\a|\b)} \ , \quad T^\a{}_{\b\c} &\equiv& 2 \Gamma^{\a}_{[\b\c]} = 2 L^\a{}_{[\b\c]} = 2 K^\a{}_{[\b\c]} \ .
\eea
We have $Q_{\a\b\c}=Q_{\a(\b\c)}$, $J_{\a\b\c}=J_{\a(\b\c)}$, and $K^\a{}_\b{}^\c=K^{[\a}{}_\b{}^{\c]}$.

We decompose the Riemann tensor into the Levi--Civita part and the distortion part as
\bea \label{Riemann}
  \!\!\!\!\!\!\!\!\!\! R^{\a}{}_{\b\c\d} &\equiv& \pat_\c \Gamma^{\a}_{\d\b}-\pat_\d\Gamma^{\a}_{\c\b} + \Gamma^{\a}_{\c\mu} \Gamma^{\mu}_{\d\b} - \Gamma^{\a}_{\d\mu}\Gamma^{\mu}_{\c\b}= \mathring R^{\a}{}_{\b\c\d} + 2 \mathring \nabla_{[\c} L^\a{}_{\d]\b} + 2 L^\a{}_{[\c|\mu|} L^\mu{}_{\d]\b} \ .
\eea
There are three independent first contractions of the Riemann tensor. We need only two: the Ricci tensor $R_{\a\b}\equiv R^{\c}{}_{\a\c\b}$ and the co-Ricci tensor $\hat R^\a{}_\b\equiv g^{\mu\nu}R^\a{}_{\mu\nu\b}$. Note that neither is symmetric. When non-metricity vanishes, we have $\hat R_{\a\b}=-R_{\a\b}$. The unique full contraction of the Riemann tensor is the Ricci scalar $R\equiv g^{\a\b}R_{\a\b}$. The Einstein tensor (defined by the variation of the Einstein--Hilbert action with respect to the metric) is $G_{\a\b}\equiv R_{(\a\b)}-\ha g_{\a\b} R$.

\subsubsection{Projective transformation} \label{sec:pro}

The projective transformation of the connection is defined as \cite{Hehl:1978}
\begin{equation} \label{projective}
	\Gamma^\c_{\a\b} \to \Gamma^\c_{\a\b} + \delta^\c{}_\b A_\a \ ,
\end{equation}
where $A_\a$ is an arbitrary vector. Under this transformation non-metricity and torsion change as
\begin{align} \label{QTpro}
		Q_{\a\b\c} &\to Q_{\a\b\c} - 2 g_{\b\c} A_{\a}  \el
		T_{\a\b\c} &\to T_{\a\b\c} - 2 g_\a{}_{[\b} A_{\c]} \ .
\end{align}
The Riemann tensor transforms as
\bea \label{Riemannpro}
  R^\a{}_{\b\c\d}\to R^\a{}_{\b\c\d}+ 2 \d^\a{}_\b \pat_{[\c} A_{\d]} \ .
\eea

If the theory is invariant under the projective transformation and either non-metricity or torsion appears only in the form \re{QTpro}, it can be removed by a projective transformation. If both non-metricity and torsion have the form \re{QTpro}, they can be completely exchanged for each other. (This is the case in $f(R)$ gravity and in the simplest Higgs inflation model \cite{Iosifidis:2018zjj, Rasanen:2018b}.) In a FLRW spacetime, due to symmetry the torsion tensor can only have terms proportional to $g_{\a[\b} u_{\c]}$ and $\epsilon_{\a\b\c\d} u^\d$, where $u^\a$ is the four-velocity orthogonal to the hypersurface of symmetry and $\epsilon_{\a\b\c\d}$ is the Levi--Civita tensor. The first term has the form \re{QTpro} and the second violates parity, so if the theory is projectively invariant and parity-invariant, it is always possible to eliminate the torsion in a FLRW spacetime. In contrast, non-metricity in a FLRW spacetime is not limited to the form \re{QTpro}, so it cannot always be transformed away.

\subsubsection{Gauss--Bonnet term}

The Gauss--Bonnet term is the second order Lovelock term \cite{Lanczos:1938sf, Lovelock:1970zsf, Lovelock:1971yv, Lovelock:1972vz}
\bea \label{GB}
  \mathcal{G} &\equiv& 6 \delta^{\a\b\m\n}_{\c\d\sigma\tau} R^{\sigma\tau}{}_{\m\n} R^{\c\d}{}_{\a\b} \el
    &=& -^{**}R_{\a\b}{}^{\c\d} R^{\a\b}{}_{\c\d} \el
    &=& R_{\a\b}{}^{\c\d} R_{\c\d}{}^{\a\b} - ( R_{\a\b} - \hat R_{\a\b} ) ( R^{\b\a} - \hat R^{\b\a} ) + R^2 \el
    &=& \mathring{\mathcal{G}} + 12 \d^{\a \b \m \n}_{\c \d \s \tau} \big( \mathring R^{\c \d}{}_{\m \n} \mathring \nabla_\a \tensor{L}{^\s _\b ^\tau} + \mathring R^{\c \d}{}_{\m \n} \tensor{L}{^\s _\a _\lambda} \tensor{L}{^\lambda _\b ^\tau} \el
    && + \mathring \nabla_\m \tensor{L}{^\c _\n ^\d} \mathring \nabla_\a \tensor{L}{^\s _\b ^\tau} + 2 \tensor{L}{^\c _\m _\lambda} \tensor{L}{^\lambda _\n ^\d} \mathring \nabla_\a \tensor{L}{^\s _\b ^\tau} + \tensor{L}{^\c _\m _\lambda} \tensor{L}{^\lambda _\n ^\d} \tensor{L}{^\s _\a _\rho} \tensor{L}{^\rho _\b ^\tau} \big) \ ,
\eea
where $\delta^{\m \n \sigma\tau}_{\a \b \c \d}\equiv\d^\m{}_{[\a} \d^\n{}_\b \d^\sigma{}_\c \d^\tau{}_{\d]}$, and $^{**}R_{\a\b\c\d}\equiv\frac{1}{4}\varepsilon_{\a\b\mu\nu}\varepsilon_{\c\d\sigma\tau} R^{\mu\nu\sigma\tau}$ is the double dual of the Riemann tensor. We have chosen to define the Gauss--Bonnet term with the Lovelock form. The definition of the Gauss--Bonnet term in the metric formulation can be extended in different inequivalent ways to the Palatini case \cite{Borunda:2008, Jim_nez_2014}. For example, the definition $-^{**}R_{\a\b}{}^{\c\d} R_{\c\d}{}^{\a\b} =-^*R_{\a\b}{}^{\c\d} {}^*R_{\c\d}{}^{\a\b}$ is equivalent to \re{GB} in the metric formulation, but not in the Palatini formulation (here $^*R_{\a\b\c\d}\equiv\frac{1}{2}\varepsilon_{\c\d\mu\nu} R_{\a\b}{}^{\mu\nu}$ is the dual Riemann tensor). The same kind of ambiguity arises for the Chern--Simons term \cite{Hassan:2024vjg}.

It is easy to see that the Gauss--Bonnet term (like the Einstein--Hilbert term) is invariant under the projective transformation \re{projective}, as the additive term in \re{Riemannpro} is symmetric in the first two indices, which are antisymmetrised in \re{GB}. This is different from the Chern--Simons term, which is not projectively invariant. Following \cite{Janssen:2019doc}, it is useful to introduce the tensor
\begin{equation} \label{Sigma}
  \Sigma^{\a\b}{}_{\c\d} \equiv 2 \delta^{\a\b\m\n}_{\c\d\sigma\tau} R^{\sigma\tau}{}_{\m\n} = 2 \delta^{\a\b\m\n}_{\c\d\sigma\tau} ( \mathring R^{\sigma\tau}{}_{\mu\nu} + 2 \mathring \nabla_{\m} \tensor{L}{^\sigma _\n ^\tau} + 2 L^\sigma{}_{\m \lambda} \tensor{L}{^\lambda _\n ^\tau} ) \ .
\end{equation}
The tensor $\Sigma_{\a\b\c\d}$ is antisymmetric in the first and the second pair of indices. The Levi--Civita part $\mathring \Sigma_{\a\b\c\d}$ is also symmetric under the interchange of the first and the second pair of indices with each other, like $\mathring R_{\a\b\c\d}$, but this is not true for $\Sigma_{\a\b\c\d}$, as it is not true for $R_{\a\b\c\d}$. The Levi--Civita part satisfies $\lcd_\a \mathring \Sigma^{\a\b}{}_{\c\d} = 0$ due to the second Bianchi identity. In terms of $\Sigma_{\a\b\c\d}$, the Gauss--Bonnet term reads simply $\mathcal{G}=3\Sigma_{\a\b}{}^{\c\d}R_{\c\d}{}^{\a\b}$. When non-metricity vanishes, $\mathcal{G}$ is a total derivative (in 4 dimensions, which we assume) \cite{Nieh:1979hf, Nieh:2007zz, Nieh:2008btw, Nieh:2018rlg}. It is also a total derivative when non-metricity has the form $Q_{\a\b\c}=B_\a g_{\b\c}$, where $B_\a$ is an arbitrary vector \cite{Babourova_1997}. (The Palatini theory where non-metricity has this form and torsion is zero is known as the metric-affine Weyl theory.) This is rather obvious, as $\mathcal{G}$ is projectively invariant and this form of non-metricity can be removed with a projective transformation. At least in spatially flat FLRW models, this is not the only form of non-metricity for which $\mathcal{G}$ is a total derivative, but $\mathcal{G}$ is not a total derivative for arbitrary non-metricity \cite{Iosifidis_2021, Janssen:2019uao}.

\subsection{Action and equations of motion}

\subsubsection{Action} \label{sec:action}

We consider a scalar field coupled to the Gauss--Bonnet term, with the action (in units where the reduced Planck mass is unity)
\begin{align} \label{action}
  S & = \int \rmd^4 x \sqrt{-g} \left[ \ha R - \ha K(\varphi) X - V(\varphi) - E(\varphi) \mathcal{G} \right] \ ,
\end{align}
where we have included the non-minimal kinetic function $K(\varphi)$ and the direct Gauss--Bonnet or Euler coupling $E(\varphi)$; we denote $X_{\a\b}\equiv\pat_\a\varphi\pat_\b\varphi$, $X\equiv X^\a{}_\a$. In the action, the Riemann tensor $R^\a{}_{\b\c\d}$ depends only on the connection, not on the metric. The action is projectively invariant. Varying the Gauss--Bonnet contribution with respect to $R^\a{}_{\b\c\d}$ gives $-6 E\Sigma^{\c\d}{}_\a{}^\b$. A direct coupling to the Ricci scalar of the form $F(\varphi) R$ could be eliminated with the conformal transformation of the metric $g_{\a\b}\to F^{-1} g_{\a\b}$, which would shift $K\to K/F$, $V\to V/F^2$, $E\to E$, so we can neglect such coupling without loss of generality.

\subsubsection{Metric equation of motion} \label{sec:metricEOM}

Varying \re{action} with respect to $g_{\a\b}$ gives the equation of motion (written in \cite{Gasperini:1992sq} using tetrads)
\bea \label{metricEOM}
  G_{\a\b} = T_{\a\b} + E ( 12 \tensor{\Sigma}{^\mu ^\nu _\sigma _( _\a}  \tensor{R}{^\sigma _\b _) _\mu _\nu} - g_{\a\b} \mathcal{G} ) \ ,
\eea
where
\bea \label{T}
  T_{\a\b} &=& K X_{\a\b} - g_{\a\b} \left( \ha K X + V \right)
\eea
 is the energy-momentum tensor. The Gauss--Bonnet contribution in \re{metricEOM} proportional to $E$ is traceless, corresponding to the fact that the Gauss--Bonnet action is conformally invariant. As the action is (as usual in the Palatini formulation) algebraic in the metric, this equation of motion involves only $E$, not its derivative. When non-metricity is zero, the Gauss--Bonnet term is a total derivative, so its contribution to the equation of motion has to vanish for $\pat_\a E=0$. This means that the term proportional to $E$ in \re{metricEOM} then has to be zero. (The Chern--Simons term, which is always a total derivative, also does not directly contribute to the metric equation of motion in the Palatini formulation \cite{Hassan:2024vjg}.) Indeed, in the metric formulation it is well-known that this term is proportional to the Lanczos--Lovelock tensor, which vanishes identically in 4 dimensions. In contrast, when non-metricity is non-zero, the Gauss--Bonnet term can contribute directly to the metric equation of motion.
 
 \subsubsection{Distortion equation of motion} \label{sec:LEOM}

Let us derive the equation for the distortion in the unconstrained case; the cases when non-metricity or torsion is put to zero a priori can be readily obtained from it by symmetry. Varying \re{action} with respect to $L_{\b\c\a}$ gives the equation of motion \cite{Gasperini:1992sq}
\bea \label{LEOM}
  && L_{\a\b\c} + L_{\c\a\b} - g_{\b\c} L_{\a\mu}{}^\mu - g_{\c\a} L^\mu{}_{\mu\b} \el
  && = - 24 \pat_\mu E \tensor{\Sigma}{^\mu_\c_\a_\b} - 24 E ( \tensor{L}{_\a ^\mu ^\nu} \tensor{\Sigma}{_\c _\mu _\b _\nu} + \tensor{L}{^\mu ^\nu _\b} \tensor{\Sigma}{_\c _\nu _\a _\mu} + \lcd_\mu \tensor{\Sigma}{^\mu _\c _\a _\b} ) \el
  && \equiv 6 E_{\a\b\c} - 24 E ( \tensor{L}{_\a ^\mu ^\nu} \tensor{\Sigma}{_\c _\mu _\b _\nu} + \tensor{L}{^\mu ^\nu _\b} \tensor{\Sigma}{_\c _\nu _\a _\mu} + \lcd_\mu \tensor{\Sigma}{^\mu _\c _\a _\b} ) \ ,
\eea
where on the second line we have defined
\bea \label{E}
  E_{\a\b\c} &\equiv& - 4 \pat_\d E \tensor{\Sigma}{^\d_\c_\a_\b} =  - 8 g_{\c\epsilon} \pat_\d E \delta^{\d\epsilon\m\n}_{\a\b\sigma\tau} ( \mathring R^{\sigma\tau}{}_{\m\n} + 2 \mathring \nabla_{\m} \tensor{L}{^\sigma _\n ^\tau} + 2 L^\sigma{}_{\m \lambda} \tensor{L}{^\lambda _\n ^\tau} ) \ .
\eea
Note that $E_{\a\b\c}=E_{[\a\b]\c}$. We define $E_\a\equiv E_{\a\mu}{}^\mu$. The Levi--Civita part of $E_{\a\b\c}$ is
\begin{align}
  & \mathring E_{\a\b\c}  = \frac{8}{9} \mathring R g_{\c[\a} \partial_{\b]} E  - \frac{4}{3} g_{\c[\a} \mathring R_{\b]}{}^\mu \partial_\m E - \frac{4}{3} \mathring R_{\c[\a} \partial_{\b]} E + \frac{4}{3} \mathring C_{\a\b\c}{}^\m \partial_\m E \ ,
\end{align}
where we have applied the decomposition of the Levi--Civita Riemann tensor into Weyl and Ricci parts:
\bea \label{Rdec}
  \mathring R_{\a\b\c\d} = \mathring C_{\a\b\c\d} + ( g_{\a[\c} \mathring R_{\d]\b} - g_{\b[\c} \mathring R_{\d]\a} ) - \frac{1}{3} g_{\a[\c}g_{\d]\b} \mathring R \ .
\eea

Because $\lcd_\mu \tensor{\mathring\Sigma}{^\mu _\c _\a _\b}=0$, the last term in the equation of motion \re{LEOM} does not contain any terms independent of $L_{\a\b\c}$. So all terms proportional to $E$ contain at least one factor of $L_{\a\b\c}$. All double covariant derivatives in \re{LEOM} are antisymmetrised, so they reduce to the Levi--Civita Riemann tensor plus terms without derivatives. Thus, while \re{LEOM} contains the divergence of $\tensor{\Sigma}{^\mu _\c _\a _\b}$, and $\tensor{\Sigma}{^\mu _\c _\a _\b}$ (defined in \re{Sigma}) involves first derivatives of $L_{\a\b\c}$, there are no true second derivatives of $L_{\a\b\c}$ in \re{LEOM}, only curvature terms. The equation of motion for $L_{\a\b\c}$ is first order, as in the Chern--Simons case. This is related to the fact that even though the Gauss--Bonnet term in the action \re{action} is quadratic in first derivatives, they are all antisymmetrised, so it is linear in the time derivative.

As the equation for $L_{\a\b\c}$ is differential instead of algebraic (like in the Chern--Simons case \cite{Hassan:2024vjg}, but unlike in the Holst and the Nieh--Yan case \cite{Langvik:2020nrs, Shaposhnikov:2020gts}), we expect there to be new degrees of freedom. This is in contrast to the metric formulation where the directly coupled Gauss--Bonnet term does not introduce any degrees of freedom and is likely stable provided the scalar field has a kinetic term \cite{Crisostomi:2017ugk}. According to the Hamiltonian analysis of \cite{Banados:2025sww}, in the unconstrained Palatini case with only the Gauss--Bonnet term (with no Einstein--Hilbert term and no scalar field), there are 9 physical degrees of freedom in regions of the solution space where the constraint matrix has maximum rank. It is not known how many are healthy and how many are ghosts.

We are interested in inflation, and will first solve the distortion equation of motion in the spatially flat FLRW universe and then use the gradient approximation to look at perturbations.

\section{Solutions} \label{sec:sol}

\subsection{Background} \label{sec:back}

\subsubsection{Solving for the distortion}

Let us solve the full connection equation of motion \re{LEOM} exactly in the spatially flat FLRW universe. The metric is
\bea \label{metricFLRW}
  \rmd s^2 = - \rmd t^2 + a(t)^2 \d_{ij} \rmd x^i \rmd x^j \ ,
\eea
where $a(t)$ is the scale factor. We start with the case when no constraints are imposed on the connection, but the FLRW solution will turn out to be identical in the zero non-metricity and the zero torsion case. Taking into account that the action is parity-invariant, by symmetry non-metricity and torsion have the form
\begin{align} \label{Qdecom}
  & Q_{\a\b\c} = q_1 (t) g_{\b\c} u_\a + q_2 (t) g_{\a ( \b} u_{\c ) } + q_3 (t) u_\a u_\b u_\c \el
  & T_{\a\b\c} = 2 t_1 (t) g_{\a [ \b} u_{\c ] } \ ,
\end{align}
where $u^\a=\d^{\a0}$, and $q_i$ and $t_1$ are functions to be determined. The projective transformation \re{projective} now reads $\Gamma^\c_{\a\b} \to \Gamma^\c_{\a\b} + A(t) \delta^\c{}_\b u_\a$, where $A(t)$ is an arbitrary function. The functions in \re{Qdecom} transform as $q_1\to q_1-2A$, $t_1\to t_1-A$; $q_2$ and $q_3$ are unchanged. Because of projective symmetry, $q_1$ and $t_1$ therefore appear only in the combination $q_1-2t_1$; it is convenient to use the variable $x\equiv q_1-2t_1+2H$. The connection equation of motion \re{LEOM} then reads
\bea
  \label{FLRW1} \!\!\!\!\!\!\!\!\!\! 0 &=& q_2 [ 1 - 2 E x ( x - 2 q_2 ) ] \\
  \label{FLRW2} \!\!\!\!\!\!\!\!\!\! 0 &=& - x + 2 H + 2 q_2 [ 2 \dot q_2 + q_2 ( - 3 x + q_2 - q_3 + 2 H ) + x^2 + x q_3 ] E - 2 x ( x - 2 q_2 ) \dot E \\
  \label{FLRW3} \!\!\!\!\!\!\!\!\!\! 0 &=& 2 x - q_3 - 4 H + [ - 8 q_2 \dot x + 4 q_2 x ( q_2 - 2 q_3 - 2 H ) + 2 x^2 q_3 ] E + 4 x ( x - 2 q_2 ) \dot E \ ,
\eea
where $H\equiv \dot a/a$, and dot denotes derivative with respect to $t$. In \cite{Iosifidis_2021} it was found that the Gauss--Bonnet term is a total derivative for $q_2=0$. However, the term proportional to $E$ in \re{FLRW3} does not vanish for $q_2=0$. There is no contradiction: imposing a constraint on the action is not equivalent to imposing it on the equation of motion. We have either $q_2\neq0$ or $q_2=0$. In the first case we can solve $q_2$ from \re{FLRW1}:
\bea
  q_2 = \ha x - \frac{1}{4 E x} \ ,
\eea
Taking the time derivative and inserting $q_2$ and $\dot q_2$ into \re{FLRW2} gives \re{FLRW3}. As there remains only one independent equation but two unknown variables $x$ and $q_3$, the system is underdetermined. Therefore $q_2\neq0$ does not lead to a physical solution. Taking $q_2=0$, equations \re{FLRW2} and \re{FLRW3} reduce to
\bea \label{LEOMFLRW}
  \label{FLRW4} 0 &=& x - 2 H + 2 x^2 \dot E \\
  \label{FLRW5} 0 &=& q_3 ( 1 - 2 x^2 E ) \ .
\eea
If $q_3$ does not vanish, it is not determined, and the remaining two equations for $x$ are overdetermined. The physical solution therefore has $q_3=0$, and we obtain $x$ from \re{FLRW4}:
\bea \label{xsol}
  x &=& \frac{ - 1+ \sqrt{1 + 16 H \dot E} }{ 4 \dot E } \ .
\eea
Note that there are no physical solutions for $16H\dot E<-1$. The distortion functions are
\bea \label{QTsolFLRW}
  q_1 - 2 t_1 &=& \frac{ - 1 - 8 H \dot E + \sqrt{1 + 16 H \dot E} }{ 4 \dot E } \simeq - 8 H^2 \dot E \el
  q_2 = q_3 &=& 0 \ ,
\eea
where the last equality on the first line is the leading order in the gradient approximation, where $\pat_\a E$ (which in this case reduces to $\dot E$) is small. With $q_2=0$, time derivatives of the connection disappear from the equation of motion, so the connection does not involve new degrees of freedom. With $q_2=q_3=0$, the terms proportional to $E$ in the equation of motion vanish, corresponding to the fact that the non-metricity that remains can be removed with a projective transformation. As noted above, for $q_2=0$ the Gauss--Bonnet term is a total derivative, so the distortion can only depend on $\pat_\a E$, not on $E$ itself, and thus vanishes when $E$ is constant. Alternatively, we can remove the torsion, which can always be done in a FLRW spacetime in a parity-invariant theory, as noted in \sec{sec:pro}. Therefore, the FLRW behaviour is identical in the unconstrained, the zero torsion, and the zero non-metricity case. This is not true for perturbations around the FLRW metric.

For a theory with the Gauss--Bonnet term alone (with no Einstein--Hilbert term and no scalar field) we have equations \re{FLRW1}--\re{FLRW3} without the linear terms and with $E=1$ (this is equivalent to the $\dot E=0$, $E\to\infty$ limit). This system of equations is underdetermined. The equations are for the unconstrained case, but putting torsion to zero a priori does not change them (it only affects the definition of $x$). If non-metricity is put to zero a priori, there is no equation of motion because the Gauss--Bonnet term reduces to a total derivative.

As the Gauss--Bonnet term is in general not a total derivative, one might have thought that it could together with the Einstein--Hilbert term lead to inflation without the need to introduce a scalar field\footnote{In this case the Gauss--Bonnet term would likely have to generate extra degree of freedom, like the $R^2$ term when added to the Einstein--Hilbert action in the metric formulation \cite{Whitt:1984pd, Hawking:1984ph, Suen:1986ip, Mijic:1986iv}.}. We have seen that this is not possible, as the Gauss--Bonnet term becomes a total derivative after solving the equation of motion for the connection. This could have been expected from the fact that in de Sitter space all rank 3 tensors are zero due to symmetry, so the distortion must vanish in de Sitter limit and be slow-roll suppressed close to it (assuming the limit is smooth).

\subsubsection{Metric action}

Inserting the spatially flat FLRW metric ansatz \re{metricFLRW} and the distortion solution \re{xsol} into the action \re{action}, we obtain an action for the metric and the scalar field alone, which looks particularly simple in terms of $x$. After some partial integrations we get (dropping boundary terms)
\begin{align} \label{actionx}
  S & = \int \rmd^4 x a^3 \left[ \frac{3}{4} x ( x - 4 H ) + \ha K \dot\varphi^2 - V + \dot E x^3 \right] \ .
\end{align}
In this form the metric and the distortion are mixed inside $x$. We want to separate the distortion contribution (which depends on the scalar field via $\dot E$) from the metric part. Inserting the definition $x=q_1-2t_1+2H$ into \re{actionx} and identifying the Levi--Civita parts of the Einstein--Hilbert and the Gauss--Bonnet term, we have (undoing some partial integrations and dropping boundary terms),
\begin{align} \label{actionFLRW}
  S & = \int \rmd^4 x a^3 \left[ \ha\mathring R + \ha K \dot\varphi^2 - V - E \mathring{\mathcal G} + \frac{3}{4} q_1^2 + \dot E q_1 ( 12 H^2  + 6 H q_1 + q_1^2 ) \right] \ ,
\end{align}
where we have used the projective symmetry to set $t_1=0$ to simplify the notation; at this point $q_1$ is simply the function of $H$ and $\dot E$ given in \re{QTsolFLRW}. We have made no approximation (apart from FLRW symmetry and spatial flatness), this is the full action for the metric and the scalar field. We now want to read off the leading order changes to the inflaton kinetic term from this action. There are no corrections to the potential, because $q_1$ vanishes when $\dot E=\dot\varphi E'$ goes to zero, where prime denotes derivative with respect to $\varphi$. This is similar to the Chern--Simons, Holst, and Nieh--Yan terms \cite{Langvik:2020nrs, Shaposhnikov:2020gts, Hassan:2024vjg}, whereas non-minimal coupling between the inflaton and gauge fields treated in the Palatini formulation can also change the potential \cite{Rasanen:2022ijc}. Considering the case when $\dot E$ is small, we insert $q_1$ from \re{QTsolFLRW} into \re{actionFLRW} and drop terms higher than quadratic in $\dot E$ to obtain
\begin{align} \label{actionFLRWgrad}
  S & = \int \rmd^4 x a^3 \left( \ha\mathring R + \ha \tilde K \dot\varphi^2 - V - E \mathring{\mathcal G} \right) \ ,
\end{align}
where
\bea \label{Ktilde}
  \tilde K \equiv K - \frac{32}{3} E'{}^2 V^2 \ ,
\eea
where we have used the metric equation of motion \re{metricEOM} to zeroth order in $\dot E$ in the spatially flat FLRW spacetime to replace $3 H^2=V$. While the distortion contribution to the action is suppressed by $\dot\varphi^2$, it can have a large impact on the kinetic term and thus on inflaton evolution even when $\dot\varphi$ is small, as $E'$ can be large. The correction in \re{Ktilde} is non-positive, so there can be (depending on the functions $K$, $E$, and $V$) a region of field space where the kinetic term is negative and the field runs uphill. However, the full kinetic term including all orders in the action \re{actionFLRW} is bounded from below for $H\dot E>0$, so the theory is not unstable in that regime. (As noted earlier, no solutions exist for $16 H \dot E<-1$.)

If slow-roll inflation is to be an attractor, the kinetic term should be positive, so for $K=1$ we must have $|E'|V<\sqrt{3/32}\approx0.3$. As we noted in \sec{sec:action}, a direct coupling of the form $F(\varphi) R$ can be removed with a conformal transformation, giving $K\to K/F$, $V\to V/F^2$, $E\to E$. In theories like Higgs inflation \cite{Bezrukov:2007, Bauer:2008} where $F$ is large and $K/F$ is small in the inflationary region, the constraint on $E'$ is thus even stronger. In theories like $\a$-attractors where $K/F$ instead grows in the inflationary region \cite{Kallosh:2013hoa}, the Palatini contribution to $\tilde K$ can be large, but it will still be important only when the kinetic term is negative or close to flipping sign. In the metric formulation with a Gauss--Bonnet term (which our FLRW case reduces to) one of the slow-roll conditions is $E'{}^2 V^2\ll K$ \cite{Satoh_2008, Mudrunka:2025xcg}, so the Palatini contribution to the kinetic term is small during slow-roll inflation, but it could flip the sign during preheating and lead to interesting phenomenology \cite{Carter:2005fu, Rashidi_2020, Mudrunka:2025xcg}.

For the Chern--Simons term in the unconstrained and the zero non-metricity case the correction to $K$ has the same form as in \re{Ktilde}, but with a positive coefficient, so there is no runaway behaviour at leading order \cite{Hassan:2024vjg}. (In the zero torsion case there is no correction, unlike for the Gauss--Bonnet term.) The correction is also non-negative for the Holst and the Nieh--Yan term \cite{Langvik:2020nrs, Shaposhnikov:2020gts}, while direct derivative coupling between the scalar field and the Riemann tensor can give a correction of either sign \cite{Germani:2010gm, Nezhad:2023dys}.

\subsection{Gradient approximation} \label{sec:pert}

\subsubsection{The expansion scheme}

As in the Chern--Simons case \cite{Hassan:2024vjg}, to find the equations for the perturbations, we solve the $L_{\a\b\c}$ equation of motion \re{LEOM} in the gradient approximation and insert the solution back into the action to obtain a theory that depends only on $g_{\a\b}$ and $\varphi$. By the gradient approximation we mean that $\pat_\a\varphi$ is small, both for time and spatial derivatives. This expansion is well suited to slow-roll inflation in the super-Hubble regime. One difference from the Chern--Simons case is that the Gauss--Bonnet term is not always a total derivative. The first term on the right-hand side of \re{LEOM} is proportional to $\pat_\a E=E'\pat_\a\varphi$ analogously to the Chern--Simons case, but the second term is proportional to $E$, and has no Chern--Simons counterpart. However, we have seen that for the spatially flat FLRW background the solution depends only on $\pat_\a E$. Therefore, in the gradient approximation even the terms proportional to $E$ are suppressed by $\pat_\a E$, as we know that they vanish when $\pat_\a E=0$. So $\pat_\a E$ works as an expansion parameter controlling the validity of our effective field theory, but unlike in the FLRW case, we can also have terms of the form $E \pat_\a E$, which are not parametrically smaller than $\pat_\a E$. The requirement that $\pat_\a E$ is small could be satisfied with a small $E'$ or $\pat_\a\varphi$: in the gradient approximation $\pat_\a\varphi$ is small, so $E'$ can be large.

Solving for the connection and inserting it back into the action to obtain a theory only for the metric and the scalar field results in higher order derivative terms that are not of the Horndeski nor DHOST form, so as a fundamental theory it would likely have ghosts \cite{Horndeski:1974wa, Langlois:2018dxi, Kobayashi:2019hrl}. (Going to higher orders in the solution for $L_{\a\b\c}$ would generate higher order derivatives with no upper limit, so the stability properties are not clear.) But we take even the original action \re{action} to be an effective theory, and apply an order reduction procedure to eliminate the higher derivatives in the action \cite{Bel:1985zz, Simon:1990ic, Simon:1990jn, Simon:1991bm, Parker:1993dk, Weinberg:2008hq, Delsate:2014hba, Solomon:2017nlh}. As in \cite{Hassan:2024vjg}, we split the Levi--Civita Riemann tensors in the action into Ricci and Weyl parts using \re{Rdec}. We then replace the Ricci tensor in the higher derivative terms (but not in the Levi--Civita Gauss--Bonnet term, which does not lead to higher order equations of motion) by using the lowest order (in the gradient expansion) metric equation of motion \re{metricEOM} with the energy-momentum tensor \re{T}, which simply reads $\mathring R_{\a\b}=V g_{\a\b}$. In the action \re{action} there is only one term linear in $L_{\a\b\c}$ that is not a boundary term, namely the second term after the last equality in \re{GB}, multiplied by $-E$. Partial integration shifts the derivative to $E$, so even for this term we need to know $L_{\a\b\c}$ only to linear order in $\pat_\a E$ to get the action to second order in $\pat_\a E$.

We consider the three cases when no constraints are imposed on the connection and when either non-metricity or torsion is taken to be zero. Let us start from the case when the connection is unconstrained.

\subsubsection{Unconstrained case}

Solving iteratively, as our ansatz we first take $L_{\a\b\c}$ in \re{LEOM} to be sourced only by $\mathring E_{\a\b\c}$ (which is proportional to $\pat_\a E$) defined in \re{E}, as all non-Levi--Civita terms in $E_{\a\b\c}$ involve more than one gradient. The solution is then
\bea \label{Lsol}
  L_{\a\b\c} &=& 6 \mathring E_{\a(\b\c)} + 3 \mathring E_{\b\c\a} + 6 \mathring E_{[\c} g_{\a]\b}  + A_\b g_{\a\c} \ ,
\eea
from which we get the disformation and the contortion,
\bea \label{JKsol}
  J_{\a\b\c} &=& 2 A_{(\b} g_{\c)\a} - A_\a  g_{\b\c} \ , \quad
  K_{\a\b\c} = 6 \mathring E_{\b [ \c\a ]} - 3 \mathring E_{\c\a\b} + ( 6 \mathring E_{[\c} - 2 A_{[\c} ) g_{\a ] \b} \ ,
\eea
as well as the non-metricity and the torsion,
\bea \label{QTsol}
  Q_{\a\b\c} &=& - 2 A_\a g_{\b\c} \ , \quad T_{\a\b\c} = 6 \tensor{\mathring E}{_\b_\c_\a} - ( 6 \mathring E_{[\b} - 2 A_{[\b} ) g_{\c ] \a} \ .
\eea
Here $A_\a$ is the arbitrary vector associated with the projective symmetry. We can choose $A_\a=\mathring E_\a$ to remove the torsion vector degree of freedom $T^\mu{}_{\a\mu}$, but it is not possible to remove torsion entirely, unlike in the FLRW case. In contrast, non-metricity has the form \re{QTpro} and is completely removed by the choice $A_\a=0$. Iterating by inserting this solution back into \re{LEOM} to evaluate the term proportional to $E$ gives zero. In the absence of non-metricity the Gauss--Bonnet term reduces to a total derivative, so there can be no terms proportional to $E$, only $E'$. We insert the solution \re{Lsol} into the action \re{action}, working up to second order in $\pat_\a\varphi$ to obtain
\bea \label{Laction}
  S &=& \int \rmd^4x \sqrt{-g} \left\{ \ha \mathring R - \ha K X - V - E \mathring{\mathcal G} + E'^2 X_{\a\b} \left[ ( 4 \mathring R^2 - 8 \mathring R_{\c\d} \mathring R^{\c\d} ) g^{\a\b} \right. \right. \el
  && \left. \left. - \frac{32}{3} \mathring R \mathring R^{\a\b} + 16 \mathring R^{\a\c} \mathring R_{\c}{}^{\b} + 32 \mathring R_{\c\d} \mathring C^{\a\c\b\d} - 64 \tensor{\mathring C}{^\a ^\c ^\d ^\e} \tensor{\mathring C}{^\b _[ _\c _\d _] _\e} \right] \right\} \ ,
\eea
There are no terms beyond second order in the Riemann tensor, as every power is accompanied by one factor of $E'$. In \cite{Gasperini:1992sq} the Riemann tensor was used as the expansion parameter, and in this case it would lead effectively to the same counting as the gradient approximation. In the zero torsion case we will discuss below in \sec{sec:zeroT}, this is not true. After the order reduction procedure, where we insert the leading order result $\mathring R_{\a\b}=V g_{\a\b}$ into \re{Laction}, the pure Ricci terms give the modification of the kinetic term we had in the FLRW case \re{actionFLRW}, the mixed Ricci--Weyl term vanishes, and we have 
\bea \label{Lactionred}
  S &=& \int \rmd^4x \sqrt{-g} \left( \ha \mathring R - \ha \tilde K X - V - E \mathring{\mathcal G} - 64 E'^2 X_{\a\b} \tensor{\mathring C}{^\a ^\c ^\d ^\e} \tensor{\mathring C}{^\b _[ _\c _\d _] _\e} \right) \ ,
\eea
where $\tilde K$ is given in \re{Ktilde}. As in the Chern--Simons case \cite{Hassan:2024vjg}, the leading Palatini correction (apart from $\tilde K$) is quadratic in the gradient of the scalar field and in the Weyl tensor.

For scalar perturbations around the spatially flat FLRW solution, the second order Weyl term contains the square of the perturbation with two spatial derivatives operating on it, for a total of four spatial derivatives, and (unlike for vector and tensor perturbations) no time derivatives. This term is therefore important only for deeply sub-Hubble modes, and can be neglected for super-Hubble modes. Due to our effective theory approach, corrections to the action for vector perturbations vanish after order reduction, and vectors remain non-dynamical (by construction of the effective theory), as in the Chern--Simons case \cite{Hassan:2024vjg}. For the tensor perturbation, we have the perturbed FLRW metric 
\bea  \label{tensormetric}
  \text{d}s^2 = -\text{d}t^2 + a(t)^2 \left( \delta_{ij} + 2 h_{ij} \right) \rmd x^i \rmd x^j \ ,
\eea
where $h^i{}_i=0$, $\partial_i h^{ij} = 0$, and spatial indices are raised and lowered with the Euclidean metric $\d_{ij}$. We insert this metric into the action \re{Lactionred}, expand to second order in $h_{ij}$, and drop the scalar field and terms linear in $h_{ij}$ to obtain
\bea
  S &=& \int \rmd^4 x a^3 \bigg\{ \ha \mathring R - E \mathring{\mathcal G} + 8 \dot E^2 \big[ a^{-4} \nabla^2 h_{ij} \nabla^2 h^{ij} + 2 a^{-2} \nabla^2 h_{ij} ( \ddot h^{ij} + H \dot h^{ij} ) \el
  && + 4 a^{-2} \dot h_{ij} \nabla^2 \dot h^{ij} + H^2 \dot h^{ij} \dot h_{ij} + \ddot h_{ij} ( \ddot h^{ij} + 2 H \dot h^{ij} ) \big] \bigg\} \ .
\eea
We use the leading order tensor equation of motion
\bea \label{tensorEOM}
  \ddot h_{ij} = - 3 H \dot h_{ij} + \frac{1}{a^2} \nabla^2 h_{ij}
\eea
 to eliminate the higher order time derivatives with the order reduction procedure, giving the action for the tensor perturbation,
\begin{align}  &
  S_{\text{tensor}} = \int \rmd^4 x a^3 \bigg\{ - \frac12 ( 1 - 8 H \dot E ) \bar g^{\a\b} \pat_\a h_{ij} \pat_\b h^{ij} - 8 H \dot E \dot h_{ij} \nabla^2 h^{ij} \el &
+ 32 \dot E^2 \big[ a^{-4} \nabla^2 h_{ij} \nabla^2 h^{ij} + a^{-2} ( \dot h_{ij} \nabla^2 \dot h^{ij} - 2 H \dot h_{ij} \nabla^2 h^{ij} )  + H^2 \dot h_{ij} \dot h^{ij} \big] \bigg\} \ ,
\end{align}
where $\bar g_{\a\b}$ is the FLRW background metric. The terms on the first line are from the Levi--Civita Ricci scalar and the Levi--Civita Gauss--Bonnet term. The second line contains the Palatini contribution. Remarkably, this term is exactly the same as in the Chern--Simons Palatini case \cite{Hassan:2024vjg}, apart from the numerical prefactor. Instead of $+32$, the Chern--Simons case has $-8$ in the unconstrained and the zero non-metricity case, and $-6$ in the zero torsion case. For the Chern--Simons term, the Palatini contribution cures the instability of one gravitational wave polarisation for large wavenumbers. Here the correction instead destabilises both polarisations. However, as in the Chern--Simons case, the validity of the effective theory based on the gradient approximation demands these corrections to be small, so within the range where the effective theory is valid, the theory remains stable.

\subsubsection{Zero non-metricity}

Let us then consider the case when non-metricity is put to zero a priori. This case (with non-zero torsion) is often called the Einstein--Cartan theory. Now there is no projective symmetry. The distortion reduces to the contortion, $L_{\a\b\c}=K_{\a\b\c}$. Unlike in the other two cases, the Gauss--Bonnet term is necessarily a total derivative, so its contribution must vanish when $\pat_\a E=0$. Hence the term proportional to $E$ vanishes in the connection equation of motion, which we obtain from \re{LEOM} by antisymmetrising over $\a$ and $\b$,
\bea \label{KEOM}
  K_{[\a\b]\c} + g_{\c[\a} K_{\b]\mu}{}^\mu = 3 E_{\a\b\c} \ ,
\eea
The leading order solution is the same as in the unconstrained case for zero non-metricity (\ie for the choice $A_\a=0$) given in \re{QTsol}. This repeats the pattern of the Chern--Simons case: there too non-metricity vanishes in the unconstrained case to leading order, so the solution agrees with the zero non-metricity case. As with the Chern--Simons term, this may not hold beyond leading order, as non-metricity and terms proportional to $E$ may appear in the unconstrained case.

\subsubsection{Zero torsion} \label{sec:zeroT}

Let us finally consider the case when torsion is taken to be zero a priori. Vanishing torsion is often taken to be part of the definition of the Palatini formulation, with the unconstrained case referred to as the metric-affine formulation. As in the zero non-metricity case, there is no projective symmetry. The distortion reduces to the disformation, $L_{\a\b\c}=J_{\a\b\c}$, whose equation of motion is \re{LEOM} symmetrised in $\a$ and $\c$. The solution at linear order in $E$ and $\pat_\a E$ is
\begin{equation} \label{Jsol}
    J_{\a\b\c} = 6 \mathring E_{\a(\b\c)} + 4 g_{\a(\b} \mathring E_{\c)} - 3 g_{\b\c} \mathring E_\a \ ,
\end{equation}
and the non-metricity is
\begin{equation}
    Q_{\a\b\c} = 6 \mathring E_{\a ( \b\c )} + 2 g_{\a(\b} \mathring E_{\c)} - 4  g_{\b\c} \mathring E_\a \ .
\end{equation}
Now the Gauss--Bonnet term is a not a total derivative, and the solution will depend not only on $\pat_\a E$, but also on $E$. We work to linear order in $\pat_\a E$ and all orders in $E$, which in the connection equation of motion means capturing all terms of order $\pat_\a E$ and $E \pat_\a E$. To do so, we insert the solution \re{Jsol} into the connection equation of motion to get a source term, working up to linear order in $\pat_\a E$. Solving for $J_{\a\b\c}$ with this source term, we obtain
\bea \label{Jsol2}
  J_{\a\b\c} &=& 6 F_{\a(\b\c)} + 4 g_{\a(\b} F_{\c)} - 3 g_{\b\c} F_\a \ ,
\eea
where
\bea
  F_{\a\b\c} &\equiv& \mathring E_{\a\b\c} + E H_{\a\b\c}{}^\d \partial_\d E \ ,
\eea
and $F_\a\equiv F_{\a\mu}{}^\mu$, with
\bea
  \!\!\!\!\!\!\!\!\!\!\!\!\!\! H_{\a\b\c\d} &\equiv& \frac {19}{27} g_{\a\d} g_{\b\c}\mathring R^2 - \frac {28}{27} g_{\a(\b} g_{\c)\d}\mathring R^2 + \frac {4}{9} g_{\d(\b}\mathring R_{\c)\a} \mathring R - \frac {40}{27} g_{\b\c}\mathring R_{\a\d} \mathring R - \frac {2}{3} g_{\a\d}\mathring R_{\b\c} \mathring R \el
  && + \frac {64}{27} g_{\a(\b}\mathring R_{\c)\d} \mathring R - \frac {5}{3} g_{\a\d} g_{\b\c} \mathring R_{\mu\nu}  \mathring R^{\mu\nu} + \frac{8}{3} g_{\a(\b} g_{\c)\d} \mathring R_{\mu\nu} \mathring R^{\mu\nu} + \frac {2}{3} g_{\a\d}\mathring R^\mu{}_\b \mathring R_{\c\mu} \el
  && + \frac {22}{9} g_{\b\c}\mathring R^\mu{}_\a \mathring R_{\d\mu} - 4 g_{\a(\b}\mathring R^\mu{}_{\c)} \mathring R_{\d\mu} + \frac {8}{9} \mathring R_{\a\d} \mathring R_{\b\c} - \frac{4}{3} \mathring R_{\a(\b} \mathring R_{\c)\d} - \frac {16}{3} \mathring R \mathring C_{\a(\b\c)\d} \el
  && + \frac{40}{9}\mathring R^\mu{}_\d \mathring C_{\a(\b\c)\mu} - \frac {16}{3}\mathring R^\mu{}_{(\b} \mathring C_{\c)\mu\a\d} - \frac{16}{3}\mathring R_{\mu\a} \mathring C_{(\b}{}^\mu{}_{\c)\d} + \frac {32}{3} \mathring R^\mu{}_{(\b} \mathring C_{\c)\d\a\mu} \el
  && - \frac {8}{3} g_{\d(\b} \mathring R^{\mu\nu} \mathring C_{\c)\nu\a\mu} + \frac {28}{3} g_{\b\c} \mathring R^{\mu\nu} \mathring C_{\a\mu\d\nu} + 4 g_{\a\d} \mathring R^{\mu\nu} \mathring C_{\mu\b\nu\c} - \frac {40}{3} g_{\a(\b} \mathring R^{\mu\nu} \mathring C_{\c)\mu\d\nu} \ .
\eea
Note that $H_{\a\b\c\d}=H_{\a(\b\c)\d}$. In the FLRW case $H_{\a\b\c\d}$ vanishes identically, in agreement with the results of \sec{sec:back}. We insert the solution \re{Jsol2} into the action \re{action}, working up to second order in $\pat_\a\varphi$. The resulting action contains 532 terms proportional to $E'{}^2$, of which 522 also involve at least one power of $E$, showing how the fact that the Gauss--Bonnet term is not a total derivative complicates the theory. We do not present the action for reasons of length. Some terms may involve a total derivative so that by partial integration they are in fact proportional to $E'^3$ and thus outside the range of validity of our approximation. Unlike in the unconstrained case, the action includes terms with up to 5 factors of the Riemann tensor, including up to 5 powers of the Ricci tensor and up to 3 powers of the Weyl tensor. So in contrast to the unconstrained (and the zero non-metricity) case, our gradient approximation is more general than the small curvature approximation used in \cite{Gasperini:1992sq}. After order reduction, the action simplifies considerably,\bea \label{actionnoT}
  S &=& \int \rmd^4x \sqrt{-g} \left[ \ha \mathring R - \ha \tilde K X - V - E \mathring{\mathcal G} - 16 f_1 E'^2 X_{\a\b} \tensor{\mathring C}{^\a ^\c^\d^\mu} \mathring C^\b{}_{(\c\d)\mu} \right. \el
  && \left. - 32 f_2 E E'^2 X_{\a\b} \tensor{\mathring C}{^\a ^\c ^\d ^\sigma} ( 3 \tensor{\mathring C}{^\b _\d ^\mu ^\nu} \tensor{\mathring C}{_\c _\mu _\sigma _\nu} - 2 \tensor{\mathring C}{^\b ^\nu _\d ^\mu} \tensor{\mathring C}{_\c _\mu _\sigma _\nu} + 2 \tensor{\mathring C}{^\b _\c ^\mu ^\nu} \tensor{\mathring C}{_\d _\mu _\sigma _\nu} ) \right] \ ,
\eea
where
\bea
  f_1 &\equiv& 1 + \frac{2}{3} E V - \frac43 E^2 V^2 + \frac{8}{27} E^3 V^3 \el
  f_2 &\equiv& 1 - \frac{4}{3} E V - \frac{4}{9} E^2 V^2 \ .
\eea
The action \re{actionnoT} differs from the action of the unconstrained and the zero non-metricity case by the third order Weyl term as well as the prefactor (and a different contraction of indices) of the second order Weyl term. As in the unconstrained case, we expand the action \re{actionnoT} to second order in the tensor perturbation $h_{ij}$, so the cubic Weyl term does not contribute, and we then reduce the order with the leading order tensor equation of motion \re{tensorEOM}. The resulting action for the tensor perturbation is
\begin{align}  &
  S_{\text{tensor}} = \int \rmd^4 x a^3 \bigg\{ - \frac12 ( 1 - 8 H \dot E ) \bar g^{\a\b} \pat_\a h_{ij} \pat_\b h^{ij} - 8 H \dot E \dot h_{ij} \nabla^2 h^{ij} \el & 
 + 24 f_1\dot E^2 \big[ a^{-4} \nabla^2 h_{ij} \nabla^2 h^{ij} + a^{-2} ( \dot h_{ij} \nabla^2 \dot h^{ij} - 2 H \dot h_{ij} \nabla^2 h^{ij} )  + H^2 \dot h_{ij} \dot h^{ij} \big] \bigg\} \ .
\end{align}
Apart from $f_1$ and the numerical prefactor, this action is the same as in the unconstrained and in the zero non-metricity case. The function $f_1$ can have either sign, so the Palatini terms may or may not lead to instability, depending on the coupling function $E$ and the potential $V$. When $EV=0$, as expected after inflation when the field settles into its minimum, the term has the same instability-inducing sign as in the unconstrained and the zero non-metricity case although, again, within the range of validity of the approximation the gradients are small and the theory is stable.

\section{Conclusion} \label{sec:conc}

We have studied the Gauss--Bonnet term directly coupled to the inflaton in the Palatini formulation, following our work on the Chern--Simons term in the Palatini formulation \cite{Hassan:2024vjg}. In contrast to the metric formulation, the Gauss--Bonnet term is not always a total derivative if non-metricity is non-zero, and in general it leads to extra degrees of freedom, some of which may be unstable. We consider three cases: when the connection is unconstrained, and when either non-metricity or torsion is put to zero a priori. We solve the connection equation of motion exactly in the spatially flat FLRW spacetime and insert it back into the action. The result is the same in all three cases. The leading order correction to the inflaton kinetic term has the same form as in the Chern--Simons case, but the sign is negative, although the full kinetic term is bounded from below. (In the Chern--Simons case the prefactor is positive in the unconstrained and the zero non-metricity case, and zero in the zero torsion case.) So the correction can only have a big impact if the kinetic term flips sign or is close to doing so, which could enable novel behaviour. The correction to the kinetic term is small in slow-roll inflation, but could play a role during preheating \cite{Carter:2005fu, Rashidi_2020, Mudrunka:2025xcg}.

We then consider perturbations. We treat the action as an effective theory and use the gradient approximation, as well as order reduction to eliminate higher derivatives. As with the Chern--Simons term, the unconstrained and the zero non-metricity case turn out to be identical at leading order, and the Gauss--Bonnet term is a total derivative, unlike in the zero torsion case. In all cases, the leading order Palatini corrections to the gravitational wave action turn out to have exactly the same form as with the Chern--Simons term, just with a different prefactor. In the Chern--Simons case, the corrections cured a metric formulation gradient instability, here they destabilise the theory further, except in the zero torsion case possibly somewhere in field space, as the prefactor is field dependent and can have either sign. However, because of the gradient approximation, the new gradient terms must be subleading, so the theory is stable within the range of validity of the approximation, as in the Chern--Simons case.

The Gauss-Bonnet term alone in the Palatini formulation in the unconstrained case has 9 degrees of freedom in the region of solution space where the constraint matrix has maximum rank \cite{Banados:2025sww}. It remains to be determined how many degrees of freedom there are, and whether any of them are ghosts, when the Einstein--Hilbert term and direct coupling to a scalar field are included, and when torsion is put to zero. If all degrees of freedom turn out to be healthy, the theory can be studied without resort to gradient approximation and order reduction to check whether the results found here have wider validity. Our analysis of the spatially flat FLRW spacetime is exact, and shows that in this case there are no new degrees of freedom.

\section*{Acknowledgments}

AH acknowledges the Mathematica xAct package \cite{xAct} used in the calculations. SR thanks Will Barker for helpful correspondence, and Laboratoire de Physique de l’Ecole normale sup\'erieure and the Astronomy Unit of Queen Mary University of London for hospitality.

\bibliographystyle{JHEP}
\bibliography{gb}

\end{document}